# A Distributed Cooperative Control Framework for Synchronized Reconnection of a Multi-Bus Microgrid

Di Shi, *Senior Member, IEEE*, Xi Chen, *Senior Member, IEEE*, Zhiwei Wang, *Member, IEEE*, Xiaohu Zhang, *Student Member, IEEE*, Zhe Yu, *Member, IEEE*, Xinan Wang, Desong Bian

*Abstract*--One critical value microgrids brings to power systems is resilience, the capability of being able to island from the main grid under certain conditions and connect back when necessary. Once islanded, a microgrid must be synchronized to the main grid before reconnection to prevent severe consequences. In general, synchronization of a single machine with the grid can be easily achieved using a synchronizer. The problem becomes more challenging when it comes to a multi-bus microgrid with multiple distributed generators (DGs) and dispersed loads. All distributed generators need to be properly controlled in a coordinated way to achieve synchronization. This paper presents a novel bi-level distributed cooperative control framework for a multi-bus microgrid. In this framework, DGs work collaboratively in a distributed manner using minimum and sparse communication. The topology of the communication network can be flexible which supports the plug-and-play feature of microgrids. Fast and deterministic synchronization can be achieved with tolerance to communication latency. Experimental results obtained from Hardware-in-the-Loop (HIL) simulation demonstrate the effectiveness of the proposed approach.

## I. INTRODUCTION

Microgrids bring resilience to power systems, which refers to the capabilities of being able to anticipate risks, limit their impacts, and bounce-back rapidly to maintain desired services through survival, adaptability, and evolution in face of consistently changing environment. The capabilities to island from and to reconnect to the main grid are considered as the critical features to this resilience concept [1]-[2]. In grid-tied mode, a microgrid exchanges power with the main grid while following the frequency and voltage set by the main grid. In case of emergency, a microgrid is disconnected from the main grid and starts to work autonomously, in a similar way to physical island, balancing its own generation and load [3]. Islanding a microgrid is an effective measure to prevent power outage and therefore to maximize the usage of renewable energy resources.

One challenging technical difficulty associated with microgrid islanding is how to resynchronize it with the main grid to prevent out-of-phase reclosing. Once islanded, a microgrid usually accelerates or decelerates due to power imbalance, losing synchronism with the main grid. When the event that triggers islanding disappears, a static switch or circuit breaker will try to connect microgrid back to the main grid. At the moment of reconnection, asynchronism can lead to severe consequences, which include system oscillations and damage to equipment, depending on how much these two systems are apart. The detrimental effects of out-of-phase reclosing are discussed in [4]-[6].

For a smooth and successful reconnection, microgrid needs to be synchronized to the main grid at the point of common coupling (PCC). Different approaches have been proposed in the literature. Generally speaking, the existing approaches can be identified into two major categories. The first type of solutions is mainly concerned with microgrids with a single DG or a single master DG [7]-[11]. The essential problem this type of solutions tries to solve is the synchronization of a single DG with the electric power system. This problem is relatively easy, and communication is only needed between the synchronization controller (synchronizer) and the (master) DG. Reference [7] summarizes the applications of various filtering algorithms and Phase-Locked Loops (PLLs). Reference [8] develops a passive monitoring scheme using PMU measurements for loss-of-mains protection for a single DG. Authors of [9] introduce a control approach that works well in system steady state but becomes incapable during system transients. Authors of [10] propose to add an auxiliary frequency error signal into the controller, which may make the reclosing time indefinite. In [11], the authors present a hybrid control approach using PMU measurements and GOOSE messages. In [12], a direct voltage and phase angle tracking algorithm is proposed.

The second type of solutions is concerned with multi-bus microgrid with multiple DGs and dispersed loads. As pointed out by [13], this problem is much more challenging than the aforementioned one as all DGs have to work cooperatively with a lot more communication needed. Without proper control protocols and coordination, DGs' actions will cancel each other out, making synchronization of the microgrid indefinite and difficult. Paper [13] proposes a centralized control architecture for microgrid synchronization, according to which each DG's control scheme has to be specifically designed. Phase tracking is achieved by adding an auxiliary input, which slows down the synchronization process. In addition, centralized control is vulnerable to single point of failure and can become cost-prohibitive when the number of DG grows substantially.

This paper furthers the work of [12] and [13] by proposing a distributed synchronization control framework, under which all DGs adjust their power outputs to cooperatively regulate voltage

This work is funded by SGCC Science and Technology Program under Grant 5455HJ160007.

D. Shi, X. Chen, Z. Wang, X. Zhang, Z. Yu, X. Wang, and D. Bian are with GEIRI North America, San Jose, CA 95134, United States. Email: {di.shi, xi.chen, zhiwei.wang, xiaohu.zhang, zhe.yu}@geirina.net.

and frequency of microgrid to track those of the main grid. As compared to the existing methods, contribution of this work can be summarized as: 1) a simple yet uniform distributed control protocol is designed which supports dynamically changing system topology and plug-and-play feature of microgrids; 2) sparse communication network can be used instead of centralized one, reducing the complexity and cost of communication infrastructure as the number of DG grows; 3) the proposed control framework is robust to communication latency and is no longer vulnerable to single point of failure; 4) direct tracking of voltage at main grid without auxiliary frequency error/offset input leads to faster and deterministic synchronization process.

The remainder of this paper is organized as follows. Section II discusses the requirements for microgrid reconnection. Section III introduces the proposed approach. Experimental results are presented in section IV while conclusions and future work are discussed in section V.

## II. BASICS FOR MICROGRID RECONNECTION

Fig. 1 shows a partial feeder multi-bus microgrid with multiple distributed generators and dispersed loads. The microgrid is connected to a main feeder via a static switch (SS), or a circuit breaker (CB), status of which determines operating state of the microgrid. When the SS/CB opens, microgrid is isolated from the utility grid, and when it closes, microgrid is reconnected.

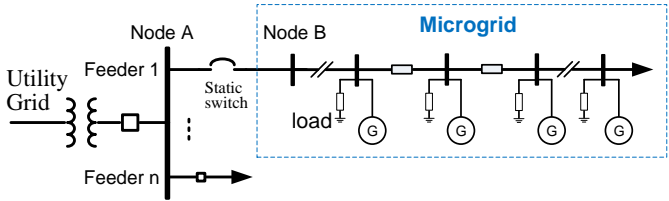

Fig. 1 A multi-bus microgrid in a typical distribution system

Microgrid islanding is usually triggered by events, for example, disturbances, blackouts, scheduled maintenance, etc. Once islanded, a microgrid will start operating in parallel with the main grid, supporting loads of its own with local distributed generation. Difference between supply and demand will cause voltage and frequency to fluctuate, losing synchronism with the main grid. When the event disappears, static switch will attempt to connect microgrid back to the main grid. A smooth transition is anticipated if voltages on both sides of the switch happen to have the same magnitude, phase angle, and frequency at the moment of reconnection. Otherwise severe consequences may be induced including huge inrush current, overvoltage, system oscillation, and equipment damage. For example, oscillation and transient power will cause excessive torsional effort which may break the shaft of a synchronous machine [14].

As shown in Fig. 1, voltage across the SS is dependent upon voltages on both side of the switch and can be evaluated by:

$$|V_{AB}| = \sqrt{|V_A|^2 + |V_B|^2 - 2|V_A||V_B|cos(\varphi_A - \varphi_B)} \quad (1)$$

where $V_A$ and $V_B$ are the voltage phasors measured at node $A$ and $B$, respectively, and $\varphi_A$ and $\varphi_B$ are the corresponding phase angles of the two phasors.

Generally speaking, voltage magnitudes at bus/node $A$ and $B$ should be regulated at close to 1 p.u. Therefore, voltage across the SS is mainly determined by angle difference $\varphi_{AB}$ at the moment of reconnection. According to (1), maximum voltage across the SS can reach 2 p.u. when the two voltages are completely out of phase. Detailed calculation of the inrush current under the worst-case scenario is discussed in [15].

The requirements for microgrid reconnection are presented in IEEE 1547™-Standard for Interconnecting Distributed Resources with Electric Power System [16]. This standard places strict requirements on the frequency difference, phase difference, and voltage magnitude difference, as summarized in Table I.

Table I Microgrid Reconnection Requirements [16]

| Avg. rating of DRs (kVA) | Freq. diff. ($\Delta f$, Hz) | Volt. mag. diff. ($\Delta V$, %) | Phase diff. ($\Delta \varphi$, degree) |
|---|---|---|---|
| 0-500 | 0.3 | 10 | 20 |
| >500-1,500 | 0.2 | 5 | 15 |
| >1,500-10,000 | 0.1 | 3 | 10 |

As noted from Table I, the tolerances become smaller as the average rating of distributed resources (DRs) goes higher. For example, as the average DR rating increases from 500 kVA to 1.5 MVA, the frequency difference tolerance decreases from 0.3 Hz to 0.1 Hz. For a microgrid with average DR rating between 1.5~10 MVA, these tolerances can be visualized as shown in Fig. 2. Selecting voltage at node $A$ as a reference, the voltage at node $B$ must lie within the shaded area and stay there for at least 277.8 milliseconds, assuming a maximum frequency difference of 0.1 Hz.

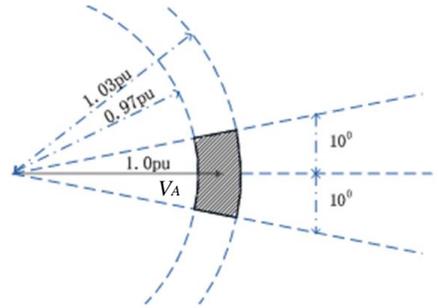

Fig. 2 Requirements for microgrid reconnection

In practice, microgrid reconnection is controlled by a synchronism checking relay, which triggers the closing of static switch once the requirements for reconnection are met.

## III. PROPOSED FRAMEWORK

This section discusses the basics of distributed cooperative control and describes the proposed framework.

### A. Graph and Distributed Control Basics

A directed graph $\mathcal{G} = \{N, \mathcal{E}\}$ with nodes $N = \{1, ..., n\}$ and edges $\mathcal{E}$ is introduced here. Each node represents an agent; each edge $(i, j)$ (pointing from $j$ to $i$) represents that information can flow from $j$ to $i$, with a weighting factor $a_{ij}$. Define neighbors of agent $i$ as $N_i = \{j \in N : (i, j) \in \mathcal{E}\}$. According to this definition agent $i$ only has access to information from its neighbors. An adjacency matrix $A = [a_{ij}] \in \mathbb{R}^{n \times n}$ associated with the graph $\mathcal{G}$ is defined as: $a_{ij} = 1$ if $(j, i) \in \mathcal{E}$, and $a_{ij} = 0$ otherwise. The Laplace matrix $L = [l_{ij}] \in \mathbb{R}^{n \times n}$ associate with the graph $\mathcal{G}$ is defined as: $l_{ij} = -a_{ij}$ when $i \neq j$, and $l_{ii} = \sum_{j=1, j \neq i}^{n} a_{ij}$. A (directed) spanning tree of $\mathcal{G}$ as a sub-graph of $\mathcal{G}$ can be defined, which is a (directed) tree that connects all the nodes in $\mathcal{G}$, as shown in Fig. 3.

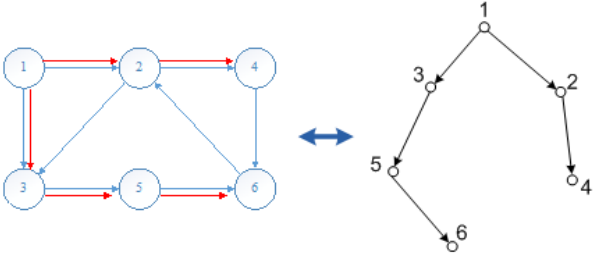

Fig. 3 A directed spanning tree of a six-node graph

Let each node be a single-state entity characterized by $\dot{x}_i = u_i$, where $u_i$ is the input as a function of agent $i$'s neighboring states $x_j$, $j \in N_i$. A consensus control problem can be formulated whose objective is to find $u_i$ so that all the state variables $x_i$'s converge to a common equilibrium point [17]. Without external input, this equilibrium will be the average value of all initial states. The practice is to adopt the following consensus protocol:

$$\dot{x}_i = u_i = \sum_{j \in N_i} a_{ij}(x_j - x_i) \qquad (2)$$

The entire system dynamic equation can be written as $\dot{x} = -Lx$ where $L$ is the Laplace matrix of the communication graph. As long as the system communication graph $\mathcal{G}$ has a spanning tree, consensus will be reached [18].

Typically the system states are required to converge to a desired external control input rather than some initial state dependent value. To achieve this, one or multiple control node(s) are introduced, which will receive a control signal $v$ from the external controller [17]. Symbol $B$ is used in the following to denote the set of control (leading) nodes. Therefore, the input signal $u_i$ can be selected accordingly, and the state function can be formulated as:

$$\dot{x}_i = u_i = b_i(v - x_i) + \sum_{j \in N_i} a_{ij}(x_j - x_i) \qquad (3)$$

where $b_i = 1$ if $i \in B$ and $b_i = 0$ otherwise. Similarly, it can be proved that as long as there exists a spanning tree of the communication graph $\mathcal{G}$, consensus will be reached and all $x_i$'s will converge to the external control signal $v$ [18].

## B. Control System Architecture

Fig. 4 shows the proposed control system architecture. Within this framework, voltage of bus $B$, $V_B$, on the microgrid side needs to be measured and compared with bus $A$ of the main grid. Fig. 4 shows the phase angles ($\varphi_A$, $\varphi_B$) and voltage magnitudes ($V_A$, $V_B$) measured at buses $A$ and $B$ are input into the controller for active synchronization control. The measuring meter can be PMU [19], or any device that can generate such measurements.

In the proposed framework, a synchronization controller sends control signals to one or several "leading" DG(s) while the remaining DGs only exchange information with its neighbor(s). With this architecture, the topology of the communication network is localized and can be very flexible. When a DG is added to the system, only one communication link between itself and its nearest neighbor needs to be added while the rest of the system including the control protocol remain intact. Compared to centralized control in which communication between the central controller and each DG is required, this proposed work requires only sparse communication between each distributed generator and its (nearest) neighbors. The topology of the communication network can be dynamic and flexible in the sense that the proposed distributed synchronization control stays effective as long as there exists a spanning tree in the communication graph after adding or removing DG(s). This control architecture supports the plug-and-play feature of microgrids.

## C. Frequency and Phase Angle Tracking

The basic idea of the proposed synchronization control strategy is to adjust voltage magnitude and frequency set points for each DG in a distributed cooperative manner so that the voltage at bus $B$ (in Fig. 4) closely follows bus $A$. Voltage and frequency adjustments are achieved via two independent control loops: 1) frequency and phase angle tracking, 2) voltage tracking.

In an islanded microgrid, all distributed generators should work under droop control so that power can be properly shared among them when load changes. The voltage droop and frequency droop are defined by the following set of equations:

$$w_i = w_i^* - k_{Pi} \cdot P_i \qquad (4)$$

$$V_i = V_i^* - k_{Qi} \cdot Q_i \qquad (5)$$

where $w_i$ and $V_i$ are the output frequency and terminal voltage of $\text{DG}_i$, $w_i^*$ and $V_i^*$ are the frequency and voltage set points of $\text{DG}_i$, $k_{Pi}$ and $k_{Qi}$ are the corresponding droop coefficients for real and reactive power.

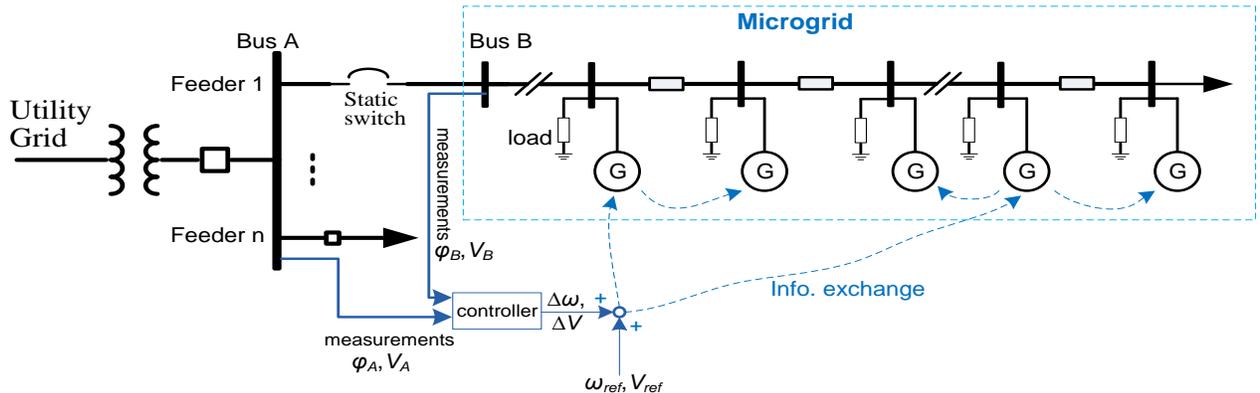

Fig. 4 The proposed distributed control architecture

In the first control loop, phase angle and frequency synchronization of the microgrid are achieved by adjusting the frequency set points of DGs in a distributed manner. A control block diagram in Fig. 5 shows how this adjustment is made at the synchronization controller level. The mode selection unit disables or enables the synchronization control. When the mode is selected to be 1, the synchronization control is activated; otherwise, it is deactivated.

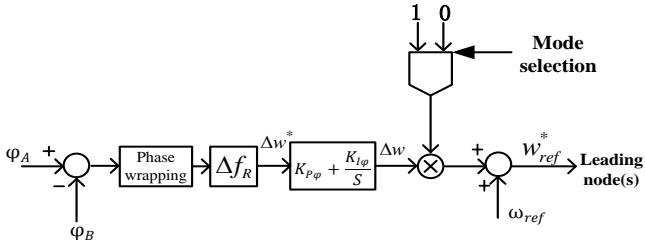

Fig. 5 Diagram for frequency and phase angle tracking

As Fig. 5 shows, phase angle difference, $\varphi_A$-$\varphi_B$ or $\Delta\varphi$, needs to be adjusted before entering the PI controller, which is due to the periodic feature of the sine wave. In this work, we propose to limit the phase angle difference $\Delta\varphi$ to the range of -180 to 180 degrees, referred to as phase wrapping. Variable $\Delta f_R$ is the reporting rate of the phase angle measurements. Output of the frequency and phase angle tracking control is a frequency adjustment signal $w_{ref}^*$, which will be sent to the leading node(s). Without loss of generality, the frequency response of each DG is modeled as a linear first-order dynamic system described by (4). Therefore, the frequency control problem is transformed into a frequency synchronization problem as discussed below.

The objective of the frequency synchronization control is to select the state variable $x_i$ and control input $u_i$ for the dynamic system described by:

$$\dot{x}_i = u_i = \sum_{j \in N_i} a_{ij}(x_j - x_i) + b_i(w_{ref}^* - x_i) \qquad (6)$$

where $b_i$=1 if $i \in B$ and $b_i$=0 otherwise.

For each DG, taking derivative on both side of the frequency droop equation (4) yields:

$$\dot{w}_i = \dot{w}_i^* - k_{Pi} \cdot \dot{P}_i \qquad (7)$$

When the frequency set point is adjusted, a DG will change its power output so that its frequency can follow the new set point. In practice we usually would like the change in each distributed generator's real power output to be proportional to its capability, that is:

$$\frac{P_1}{P_{1\_max}} = \frac{P_2}{P_{2\_max}} = \cdots = \frac{P_i}{P_{i\_max}} \qquad (8)$$

where $P_{i\_max}$ is the maximum real power generation capability of $\text{DG}_i$.

Equation (8) is equivalent to (9) since the droop coefficient $k_{Pi}$ is, as common practice, selected based on each unit's maximum real power generation capability.

$$k_{P1}P_1 = k_{P2}P_2 = \cdots = k_{Pi}P_i \qquad (9)$$

It should be noted that although in this work droop coefficients are chosen based on DG capacities, there can be other ways of selecting these coefficients. For example, reference [20] discusses the selection of droop coefficients from system stability point of view.

Based on equation (4), select the state variable to be each DG's frequency set point:

$$\dot{w}_i^* = \dot{w}_i + k_{Pi} \cdot \dot{P}_i \qquad (10)$$

The frequency synchronization problem is therefore transformed into a synchronization problem for the following linear first-order multi-agent dynamic system:

$$\begin{cases} \dot{w}_1 + k_{P1} \cdot \dot{P}_1 = u_{w1} \\ \dot{w}_2 + k_{P2} \cdot \dot{P}_2 = u_{w2} \\ \quad\vdots \\ \dot{w}_i + k_{Pi} \cdot \dot{P}_i = u_{wi} \\ \quad\vdots \end{cases} \qquad (11)$$

The controller input of $\text{DG}_i$ can be selected as:

$$u_{wi} = \sum_{j \in N_i} \left[ a_{ij}(w_j - w_i) + a_{ij}(k_{Pj}P_j - k_{Pi}P_i) \right] + b_i(w_{ref}^* \\ - w_i) \qquad (12)$$

According to (12), the frequency and phase angle tracking control loop for $DG_i$ can be visualized as shown in Fig. 6.

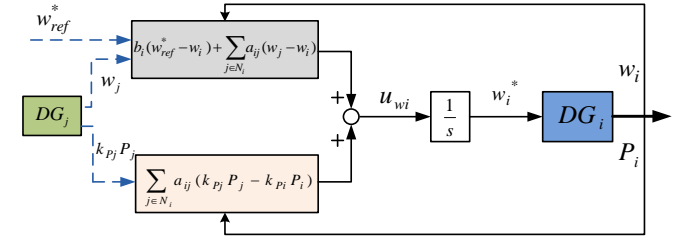

Fig. 6 Frequency and phase angle tracking control for $\text{DG}_i$

Therefore, for frequency and phase angle tracking, each DG only needs frequency(s) and power output(s) from its neighbor(s). As compared to the traditional centralized control structure, the control signal (output in Fig. 5) can be sent to only one leading DG and the rest DGs will follow the leading node based on the communication digraph.

### D. Voltage Tracking

The voltage tracking loop is responsible for regulating the microgrid voltage at PCC to match the main grid, by adjusting the voltage set points of all distributed generators. Within the synchronization controller, a diagram for calculating this voltage set point adjustment is shown in Fig. 7. The mode selection unit disables/enables the voltage tracking. Similiarly, when the mode is selected to 1, tracking is activated; otherwise, it is deactivated.

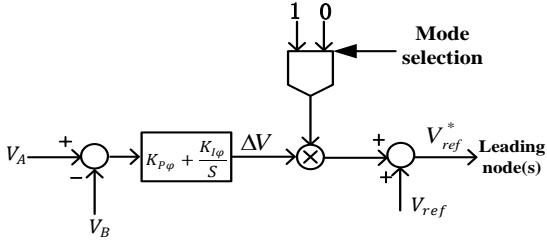

**Fig. 7** Control block diagram for voltage set point adjustment

The objective of the voltage tracking problem is to identify the state variable $x_i$ and control input $u_{vi}$ for the following first-order dynamic system:

$$\dot{x}_i = u_i = \sum_{j \in N_i} a_{ij}(x_j - x_i) + b_i(V_{ref}^* - x_i) \quad (13)$$

where $b_i=1$ if $i \in B$ and $b_i=0$ otherwise.

For each DG, the droop control function is defined in (5). Taking derivative on both sides of (5) yields:

$$\dot{V}_i = \dot{V}_i^* - k_{Qi} \cdot \dot{Q}_i \quad (14)$$

When the voltage set point is adjusted, a DG will change its reactive power output so that its terminal voltage can follow the new set point. In practice, we usually would like the reactive power sharing among DGs to follow their capacities:

$$\frac{Q_1}{Q_{1\_max}} = \frac{Q_2}{Q_{2\_max}} = \cdots = \frac{Q_i}{Q_{i\_max}} \quad (15)$$

where $Q_{i\_max}$ is the reactive power capability of DG$_i$.

Equations (15) is equivalent to (16) as the voltage droop coefficients are, as common practice, selected based on DG's reactive power capabilities:

$$k_{Q1}Q_1 = k_{Q2}Q_2 = \cdots = k_{Qi}Q_i \quad (16)$$

Again, it should be noted that although in this work droop coefficients are chosen based on DG capacities, there can be other ways of selecting them. For example, authors of [20] discuss the selection of droop coefficients from system stability point of view.

According to (5), select state variable of the aforementioned first-order dynamic system to be the voltage set point of each DG. For DG$_i$, the following equation can be written:

$$\dot{x}_i = \dot{V}_i^* = \dot{V}_i + k_{Qi} \cdot \dot{Q}_i = u_{Vi} \quad (17)$$

Therefore, the voltage tracking problem can be transformed into a synchronization problem for the following linear first-order multi-agent system:

$$\begin{cases} \dot{V}_1 + k_{Q1} \cdot \dot{Q}_1 = u_{V1} \\ \dot{V}_2 + k_{Q2} \cdot \dot{Q}_2 = u_{V2} \\ \vdots \\ \dot{V}_i + k_{Qi} \cdot \dot{Q}_i = u_{Vi} \end{cases} \quad (18)$$

According to (13), the control input for DG$_i$ is designed as:

$$u_{Vi} = \sum_{j \in N_i} [a_{ij}(V_j - V) + a_{ij}(k_{Qj}Q_j - k_{Qi}Q_i)] + b_i(V_{ref}^* - V_i) \quad (19)$$

According to (19), the voltage tracking control loop for DG$_i$ can be visualized as shown in Fig. 8.

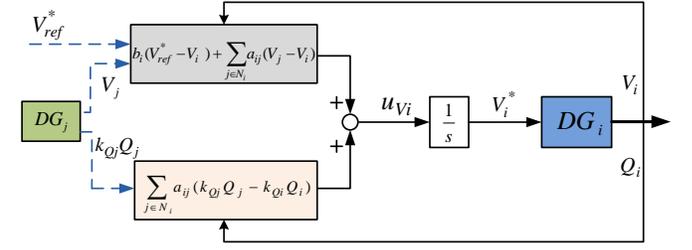

**Fig. 8** Voltage tracking control for DG$_i$

Therefore, for voltage tracking, each DG only needs voltage manitude(s) and reactive power output(s) from its neighbor(s). The synchronization controller sends a voltage reference signal $V_{ref}^*$ to the leading node(s) and the remaining DGs will follow the leading node(s) according to the communication digraph.

As the proposed synchronization control is built upon the droop control loops, it does not affect any of the existing functions (e.g., load sharing) of the DGs.

## IV. Experimental Validation

As shown in Fig. 9(a), a five-bus microgrid with four DGs and dispersed loads is employed to validate the proposed approach. Bus 5 is connected to the distribution system (main grid) through a static switch. The communication digraph is shown in Fig. 9(b). This microgrid has been implemented in the real-time digital simulator (OPAL-RT OP5600), and the controller is carried out in an external laptop as shown in Fig. 10 [21]. All four DGs are modeled as power electronic interfaced sources as discussed in [18].

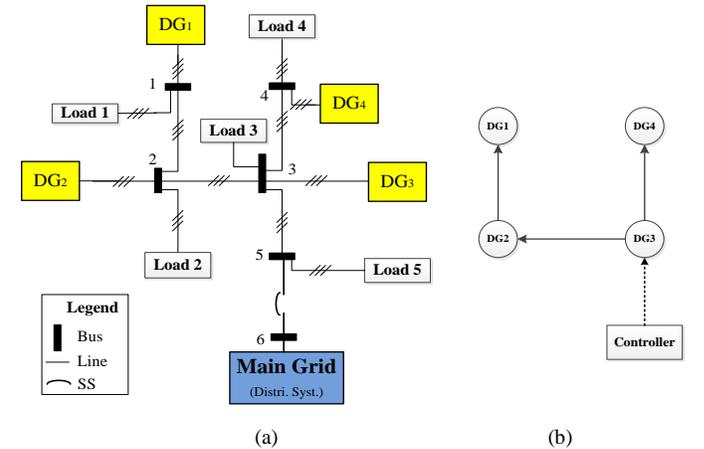

**Fig. 9** (a) 5-bus microgrid test bed  (b) communication digraph

As Fig. 10 shows, the experimental validation is conducted using HIL simulation. The electrical parts of the testbed, including the main grid, microgrid, dispersed loads, and distributed generators are modeled in Matlab/Simulink in Opal-RT. Voltage and frequency tracking loops for each individual distributed generator (shown in Fig. 6 and Fig. 8) are modeled in Simulink as well. The synchronization controller is coded in C/C++ and implemented in an external laptop. Measurements (voltages and frequencies) at bus 5 & 6 are sent out via the front

IOs of OP5600 using Ethernet communication to the synchronization controller, out of which the control commands ($w_{ref}^*$ and $V_{ref}^*$) are sent back to the simulator through Ethernet communication as well.

The microgrid test bed is briefly described below:

- System voltage: 3-Phase, 208V (L-L), 60Hz.
- Line impedances are given in Table II. The R/X ratios of lines are not uniform and range from 1 to 3.
- Load information is given in Table III.
- The ratio of DGs' capacities (DG1: DG2: DG3: DG4) is 1: 2: 3: 4, and droop coefficients of DGs are shown in Table IV.

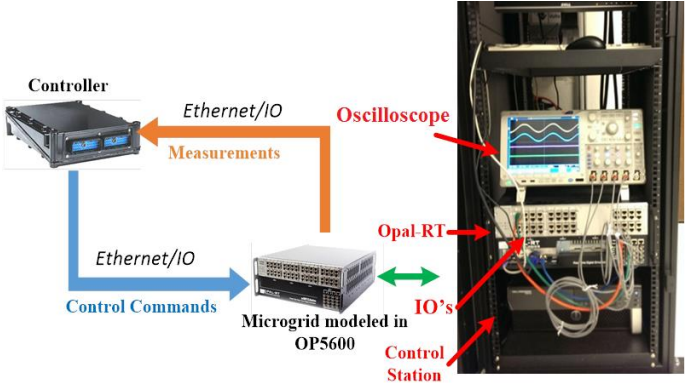

Fig. 10 System setup for the Hardware-in-the-Loop Simulation

TABLE II LINE PARAMETERS

| From | To | Impedance (Ohm per km) | | Length (feet) |
| | | Resistance | Reactance | |
|------|-----|------------|-----------|---------------|
| 1 | 2 | 0.52 | 0.37 | 500 |
| 2 | 3 | 0.29 | 0.13 | 2000 |
| 3 | 4 | 0.78 | 0.42 | 1000 |
| 3 | 5 | 0.19 | 0.12 | 100 |

TABLE III LOAD INFORMATION

| Bus | Load | |
| | Active (kW) | Reactive (kVar) |
|-----|-----------------|------------------|
| 1 | 10 | 10 |
| 2 | 5 | 2 |
| 3 | 3 | 2 |
| 4 | 2 | 2 |
| 5 | 3 (+10 @t=13) | 3 (+5 @t=13) |

TABLE IV GENERAL DROOP COEFFICIENTS

| DG | Frequency Droop $D_{p,i}$ (rad/kW) | Voltage Droop $D_{q,i}$ (V/kVar) |
|-----|-----------------|------------------|
| 1 | $4\pi$ | 0.4 |
| 2 | $2\pi$ | 0.2 |
| 3 | $4\pi/3$ | 0.13 |
| 4 | $\pi$ | 0.10 |

## A. Case Study I: Approach Validation

In this study, microgrid operates in islanded mode throughout the simulation. The objective is to see whether and how fast the aforementioned reconnection requirements are met using the proposed control. Initially, frequency of microgrid stabilizes at 59.9 Hz, and voltage at bus 5 stabilizes at 0.975 p.u., while frequency at the main grid is around 60 Hz and voltage is about

1.01 p.u. (measured at bus 6). The proposed control is enabled at $t$=6.0 second (by setting *mode* selection to 1). Simulation results are presented in Fig. 11-Fig. 13.

Fig. 11 shows the frequencies and voltage magnitudes at bus 5 (main grid side) and 6 (microgrid side), respectively. When control is enabled, frequency and voltage of microgrid start to closely track the main grid in about 5 seconds.

Fig. 12 shows the phase angle difference between the microgrid and the main grid, and the voltage across the SS. Before enabling the control, the phase angle difference varies between -180 to 180 degrees; at $t$=6 sec, it reaches 89 degrees; 3 seconds after control is enabled, it drops to well below 2 degrees. The voltage across the SS varies between 0 and 2 p.u. before control is applied and stays well below 0.02 p.u. after applying the proposed control.

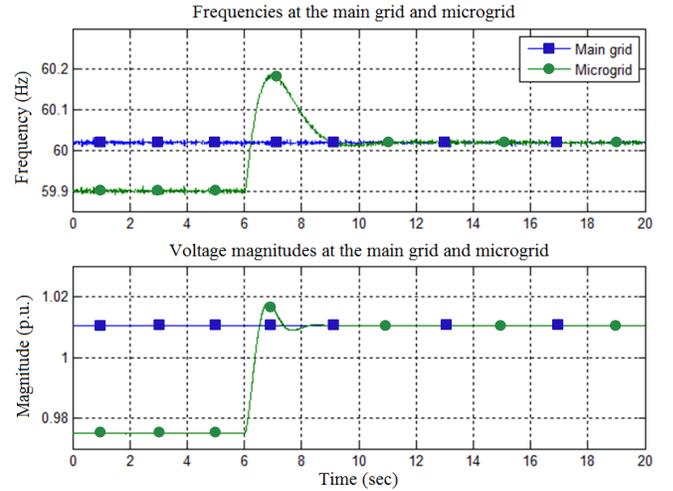

Fig. 11 Frequency and voltage on both sides of the SS (case I)

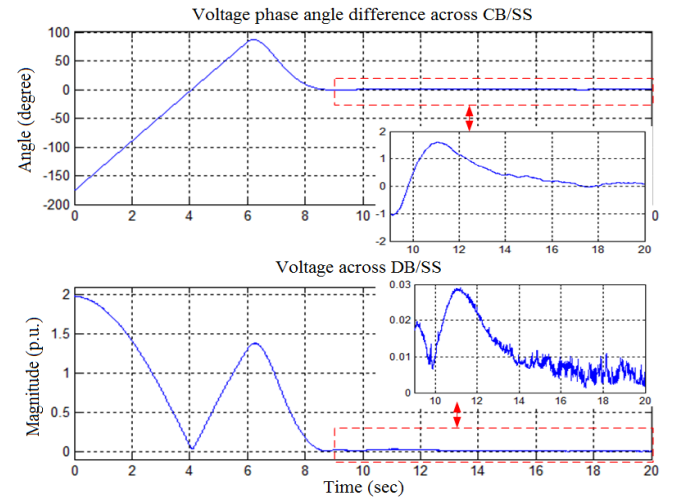

Fig. 12 Voltage and phase angle differences across the SS (case I)

As Fig. 13 shows, both real and reactive power sharings among DGs follow the ratio of their capacities (Table I). The aforementioned microgrid reconnection criteria (Table I) is satisfied with the proposed distributed control, and therefore the microgrid can be safely and smoothly reconnected to the main grid any time after the 9[th] second.

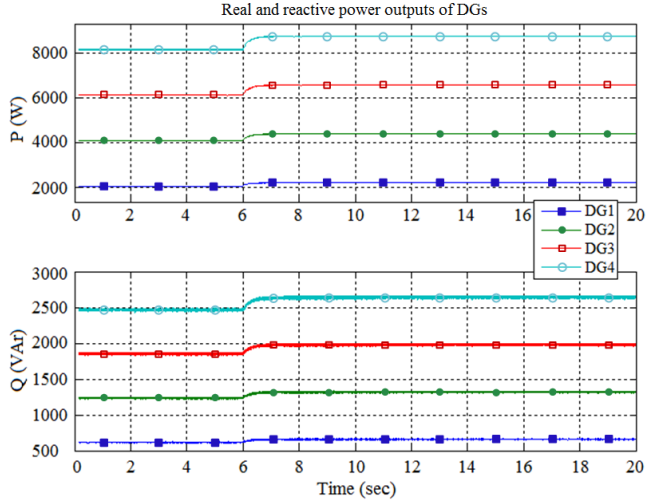

Fig. 13 Variations in outputs of DGs (case I)

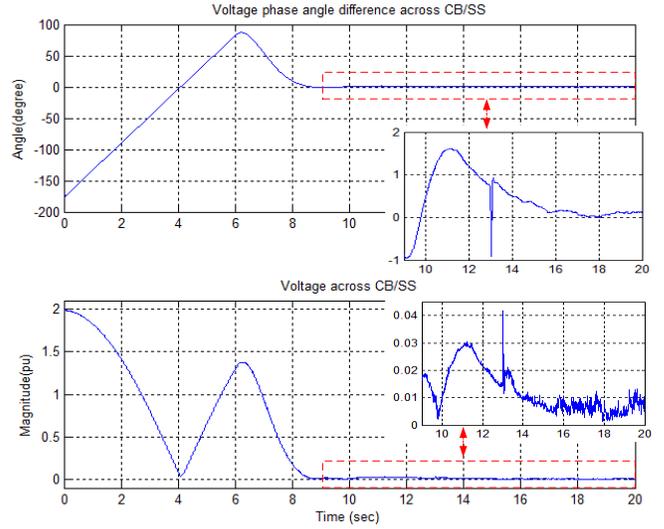

Fig. 15 Voltage and phase angle differences across SS (case II)

## B. Case Study II: Influence of Load Variations

In the second study, a considerable load increase is observed, during the synchronization process, at bus 5 (10kW+5kVAr at $t$=13s). The objective is to test the robustness of the proposed control under system disturbance. The initial system conditions are the same as case I. When the control is enabled at $t$=6 sec., voltage and frequency of microgrid start to closely track those of the main grid in about 5 seconds. Simulation results are shown in Fig. 14-Fig. 16.

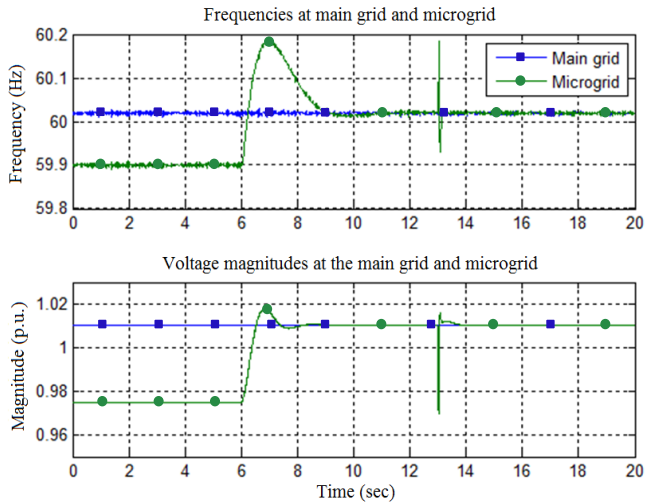

Fig. 14 Frequency and voltage on both sides of the SS (case II)

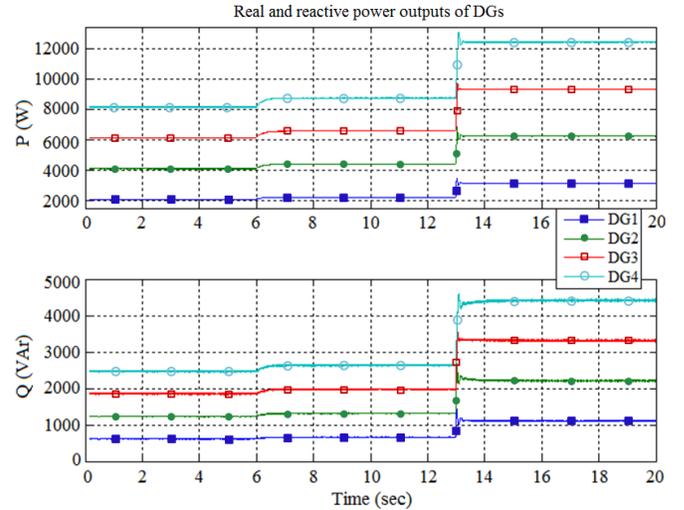

Fig. 16 Variations in outputs of the four DGs (case II)

## C. Case Study III: Influence of Communication Delay

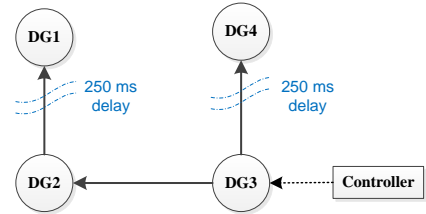

Fig. 17 Communication delay between distributed generators

As Fig. 14-Fig. 15 show, the step increase in load causes microgrid frequency and voltage to deviate from the main grid, which leads to increase of the voltage across the SS. However, due to the synchronization control, influences of this disturbance vanish quickly, and the microgrid remains synchronized with the main grid. Reconnection can occur at any time 0.5 second after the disturbance. Another factor for the fast recovery of the system was that the four DGs considered in the simulation are power electronic interfaced DGs which act very quickly. Fig. 16 shows both real and reactive power sharings among DGs follow the ratio of their capacities.

In the third study, to verify the robustness of the proposed distributed control, communication latency/delay between DGs is considered. Considering its scope, communication network used for a microgrid or among field devices usually belongs to NAN ( Neighborhood Area Network) in which WiMAX, LTE/4G can be implemented [22]. The typical point to point delay for WiMAX, LTE/4G, or etc. is on the order of milliseconds to tens of milliseconds [23]-[26]. This case study considers the worst-case scenario, in which communication between DG1 & 2, and DG3 & 4 are continuously delayed by 250 *ms* throughout the simulation, as shown in Fig. 17. The rest of the simulation conditions are the same as those in case I.

Experimental results are shown in Fig. 18-Fig. 19, in which the results from this case (case III) (referred to as the 'with delay' case) are compared against case 1 (referred to as the 'no delay' case). Basically, the comparison reveals that although discrepancy is observed between the cases with and without communication delays, the synchronization process is not affected much. Therefore, the proposed distributed control framework is robust to communication latency/delay.

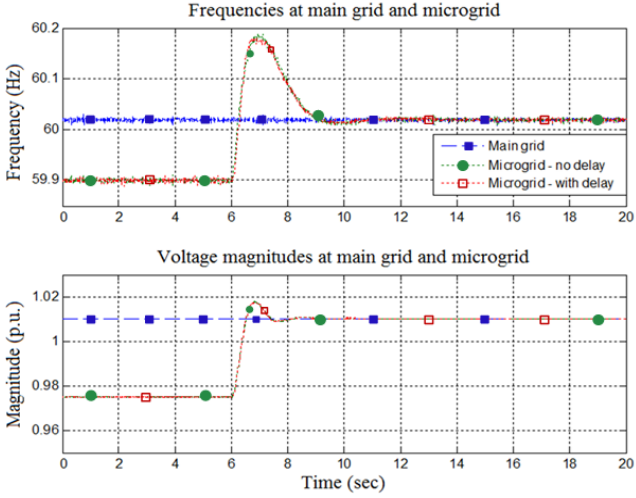

Fig. 18 Frequency and voltage on both sides of the SS (case III)

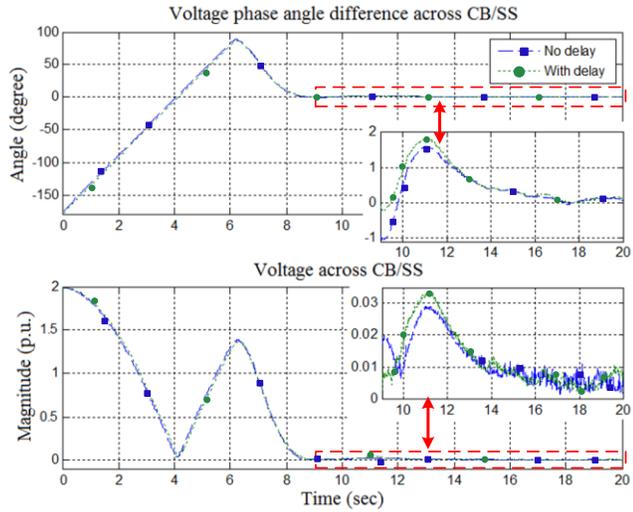

Fig. 19 Voltage and phase angle differences across the SS (case III)

### D. Case Study IV: Influence of DG Inertia and Communication Delay

In the fourth study, both inertia of distributed generators and delay in communications are considered in the simulation to further test the robustness of the proposed control. Models of all four DGs are modified by adding a virtual internia loop based on approaches discussed in [27] and [28]. The same communication delays of 250 ms between DG1 &2, and DG3&4 are considered throughout the simulation. The rest simulation conditions are the same as those in case I.

Experimental results are summarized as shown in Fig. 20-Fig. 21, in which the results from this case study (referred to as the 'with inertia, with delay' case) are compared against case 1 (referred to as the 'no inertia, no delay' case). Basically, the comparison reveals that with communication delay and inertia of DGs considered, the synchronization process takes longer, under the worst case scenario in terms of communication delay, from roughly 3 seconds (in case I) to 7 seconds. After synchronization control is enabled, frequency and voltage of the microgrid start to follow the main grid in a slower speed as compared to case I. In addition, less overshoots in frequency and voltage magnitude are observed. Therefore, the proposed distributed control framework is robust to communication delay when inertia is considered for all four DGs.

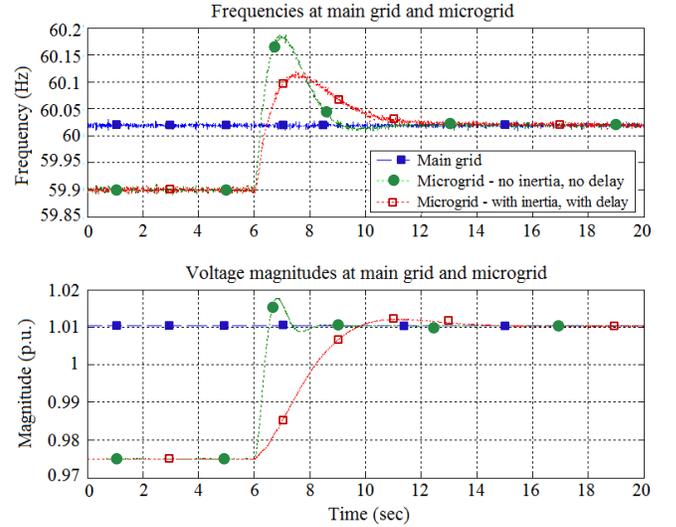

Fig. 20 Frequency and voltage on both sides of the SS (case IV)

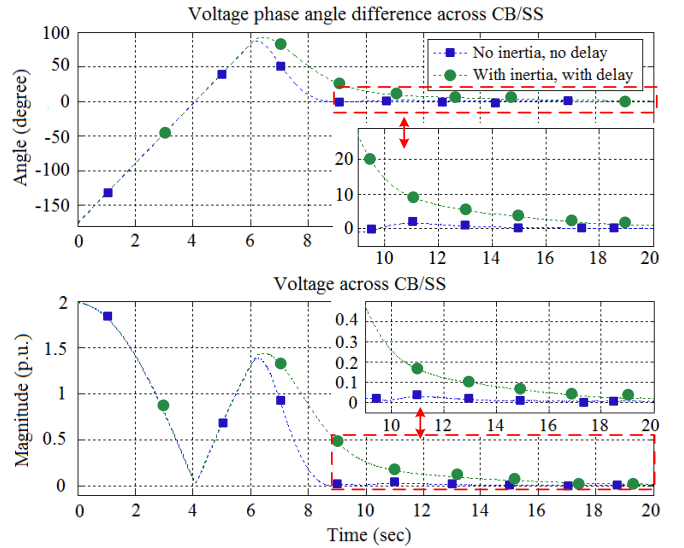

Fig. 21 Voltage and phase angle differences across the SS (case IV)

### V. CONCLUSION

A distributed cooperative control framework is proposed for synchronized reconnection of a multi-bus microgrid with multiple distributed generators and dispersed loads. Using the proposed approach, all distributed generators work in a cooperative way to regulate the frequency and voltage of the microgrid and to closely track those of the main grid. The proposed approach greatly enhances the resilience of microgrid by eliminating the possibility of out-of-phase reclosing. Phase angle measurements are directly used for frequency and phase angle tracking without the need of auxiliary input(s), which results in faster and deterministic synchronization process. The proposed control uses sparse communication infrastructure and is adaptive to network topological changes, which supports the plug-and-play feature of microgrid. The proposed framework is also robust to latency in the communication network. Experimental results obtained from a hardware-in-the-loop (HIL) testbed demonstrate the effectiveness of the proposed approach.

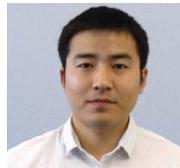

**Di Shi** (M'12, SM'17) received the B.S. degree from Xi'an Jiaotong University, Xi'an, China, in 2007, and the M.S. and Ph.D. degrees from Arizona State University, Tempe, AZ, USA, in 2009 and 2012, all in electrical engineering. He currently leads the PMU & System Analytics Group at GEIRI North America, San Jose, CA, USA. Prior to that, he was a researcher at NEC Laboratories America, Cupertino, CA, and Electric Power Research Institute (EPRI), Palo Alto, CA. He served as Senior/Principal Consultant for eMIT, LLC. and RM Energy Marketing, LLC. He has published over 60 journal and conference papers and hold 14 US patents/patent applications. One Energy Management and Control (EMC) technology he developed has been successfully commercialized. He received the Best Paper Award at 2017 IEEE PES General Meeting.

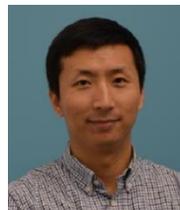

**Xi Chen** (S'07-M'13-SM'16) received MSc and PhD from King's College London, University of London, and the Hong Kong Polytechnic University, in 2005 and 2008, respectively. He was a postdoc fellow at Institute of Software, Chinese Academy of Science, from 2011 to 2013, and Research Associate at the Hong Kong Polytechnic University in 2009. He joined State Grid Information and Telecommunication Company in 2009. Currently, he is the director of Department of Development and Planning at GEIRI North America, San Jose, USA. His research interests include Internet of Things, smart grid, electric vehicle charging infrastructure, nonlinear system, complex network analysis and its applications.

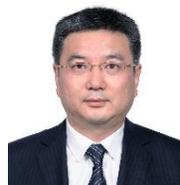

**Zhiwei Wang** (M'16) received both B.S. and M.S. degrees in electrical engineering from Southeast University, Nanjing, China, in 1988 and 1991, respectively. He is President of GEIRI North America, Santa Clara, CA, USA. His research interests include power system operation and control, relay protection, power system planning, and WAMS.

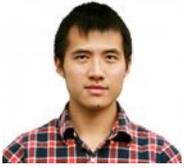

**Xiaohu Zhang** (S'12) received the B.S. degree in electrical engineering from Huazhong University of Science and Technology, Wuhan, China, in 2009, the M.S. degree in electrical engineering from Royal Institute of Technology, Stockholm, Sweden, in 2011, and the Ph.D. degree in electrical engineering at The University of Tennessee, Knoxville, in 2017.

Currently, he works as a power system engineer at GEIRI North America, Santa Clara, CA, USA. His research interests are power system operation, planning and stability analysis.

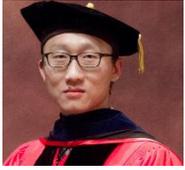

**Zhe Yu** (S'11, M'17) received his B.E. degree from Department of Electrical Engineering, Tsinghua University, Beijing, China in 2009, M.S. degree from Department of Electrical and Computer Engineering, Carnegie Mellon University, Pittsburgh, PA, USA in 2010, and Ph.D. from the School of Electrical and Computer Engineering, Cornell University, Ithaca, NY, USA in 2016. He joined GEIRI North America in 2017. His current research interests focus on the power system and smart grid, demand response, dynamic programming, and optimization.

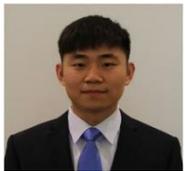

**Xinan Wang** (S'15) received the B.S. degree in electrical engineering from Northwestern Polytechnical University, Xi'an, China, in 2013, and the M.S. degree in electrical engineering from Arizona State University, Tempe, AZ, USA, in 2016. He currently works as a research assistant in the Advanced Power System Analytics Group at GEIRI North America. His research interests include WAMS and grid integration of renewables.

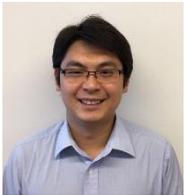

**Desong Bian** received his B.S. degree from Department of Electrical and Computer Engineering, Tongji University, Shanghai, China in 2007, M.S. degree from Department of Electrical and Computer Engineering, University of Florida, Gainesville, FL, USA in 2011, and Ph.D. from the School of Electrical and Computer Engineering, Virginia Tech, Arlington, VA, USA in 2016. He joined GEIRI North America in 2017. His research interests include PMU related applications, demand response, communication network for smart grid, etc.